\documentclass[aps,prl,twocolumn,amsmath,amssymb,superscriptaddress,floatfix]{revtex4-2}

\usepackage{graphicx}
\usepackage{bm}
\usepackage{xcolor}
\usepackage[colorlinks=true,citecolor=blue!60!black,linkcolor=blue!60!black,urlcolor=blue!60!black]{hyperref}

\newcommand{\br}{\mathbf r}
\newcommand{\bk}{\mathbf k}
\newcommand{\bq}{\mathbf q}

\newcommand{\bQ}{\mathbf Q}
\newcommand{\bPi}{\boldsymbol\Pi}
\newcommand{\cd}{\hat c^\dagger}
\newcommand{\ca}{\hat c}
\newcommand{\dd}{\hat d^\dagger}
\newcommand{\da}{\hat d}
\newcommand{\FS}[1]{|{\rm FS}_{#1}\rangle}
\newcommand{\EFS}[1]{E_{\rm FS}^{#1}}

\begin{document}

\title{Phase diagram of the attractive lattice Fermi polaron and the instability of Nagaoka ferromagnetism}

\author{Gerard Pascual}
\affiliation{Coll\`ege de France, 11 place Marcelin Berthelot, 75005 Paris, France}
\author{Kris Van Houcke}
\email{kris.van.houcke@phys.ens.fr}
\affiliation{Laboratoire de Physique de l'Ecole Normale Sup\'erieure, ENS, Universit\'e PSL, CNRS, Sorbonne Universit\'e, Universit\'e Paris Cit\'e, F-75005 Paris, France}

\date{\today}

\begin{abstract}
On the square lattice, the attractive polaron competes not only with the dimeron but also with Yang's $\eta$ dimer, an exact eigenstate that maps onto the Nagaoka ferromagnet. With two-particle--hole Ans\"atze benchmarked by diagrammatic Monte Carlo, we find that the ground state evolves from polaron to dimeron to $\eta$ dimer at low filling, and, beyond a triple point, directly from polaron to $\eta$ dimer. Above 25\% filling it stays a polaron. As the $\eta$ dimer is exact, its boundary is a rigorous instability criterion for Nagaoka ferromagnetism, sharper than the previous best and logarithmic at low filling.
\end{abstract}

\maketitle

\emph{Introduction.}---When a single distinguishable particle is added to a Fermi sea and interacts with the particles of the sea,  it drags along a cloud of particle--hole excitations and becomes a quasiparticle with renormalized properties, the Fermi polaron~\cite{Massignan2014,Massignan2026}. 
Since its first observations in spin-imbalanced Fermi gases~\cite{Shin2008Nature,Shin2008,Schirotzek2009,Nascimbene2009,Nascimbene2010}, this paradigmatic quasiparticle has been studied extensively. 
In the continuum, a simple variational Ansatz containing a single particle--hole excitation~\cite{Chevy2006} gives the energy of the attractive Fermi polaron surprisingly accurately, because the contributions of higher-order excitations largely cancel~\cite{Combescot2007,Combescot2008}. As the attraction increases, the ground state changes from the polaron to a dimeron, a bound dimer dressed by the Fermi sea, both in three~\cite{Prokofev2008,Punk2009,Mora2009,Vlietinck2013} and in two dimensions~\cite{Parish2011,Parish2013,Vlietinck2014}.

On a lattice the problem changes qualitatively, because the bandwidth is finite and Pauli blocking grows with the filling. In the continuum, a sufficiently strongly attracted impurity always ends up as a bound dimer. On a lattice, however, this is no longer guaranteed. In Ref.~\cite{Pascual2025}, using variational Ans\"atze with a single particle--hole excitation and diagrammatic Monte Carlo (DiagMC) 
we found evidence that the transition from the polaron to the dimeron disappears for majority fillings $\rho_\uparrow$ above about $0.2$. 
The nature of the ground state at strong coupling remained open, and an accurate ground-state phase diagram as a function of filling and coupling has not been established.
Experimentally, quantum-gas microscopes now give site-resolved access to the attractive Hubbard model~\cite{Mitra2018,Hartke2023} and to single dopants in magnetic backgrounds~\cite{Koepsell2019,Xu2023,Lebrat2024,Prichard2024}, so that these questions might be within experimental reach.

\begin{figure}[t]
\includegraphics[width=\columnwidth]{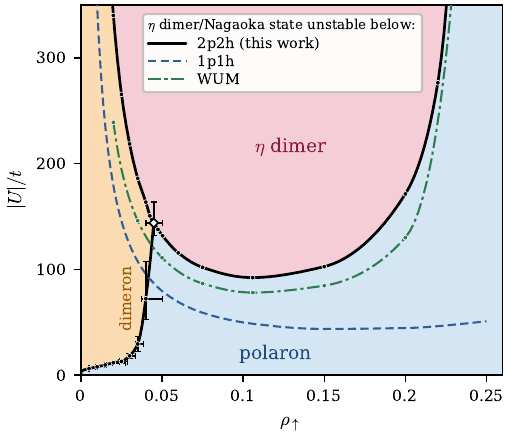}
\caption{Ground-state phase diagram of a spin-down fermion in a spin-up Fermi sea on the square lattice, in the thermodynamic limit, from 2p2h variational Ansatz. 
Solid line bounding the $\eta$-dimer region: the 2p2h polaron (or dimeron) energy crosses the exact $\eta$-dimer energy $U-\epsilon_F$. Below this line the $\eta$ dimer is not the ground state, i.e.\ the Nagaoka ferromagnet of the repulsive model is unstable against a single spin flip. Solid line between the polaron and dimeron regions: crossing of the two 2p2h energies. Below $\rho_\uparrow=0.005$ this boundary is extrapolated to the continuum limit (see text).
Dashed: same criterion with 1p1h variational Ansatz~\cite{Chevy2006,Pascual2025}. Dash-dotted: the instability criterion of Wurth, Uhrig and M\"uller-Hartmann (WUM)~\cite{Wurth1996}, the most stringent previous rigorous criterion on the square lattice.
Error bars on the polaron--dimeron line, and on the triple point (diamond), mark where the polaron--dimeron energy difference is below the $10^{-3}t$ resolution of our thermodynamic-limit quadrature.
}
\label{fig:phase}
\end{figure}

\begin{figure}[t]
\includegraphics[width=\columnwidth]{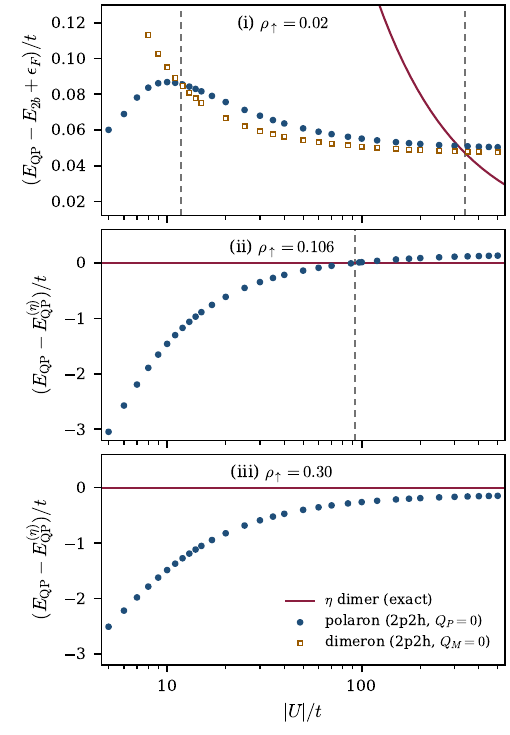}
\caption{The three scenarios of Fig.~\ref{fig:phase}, along cuts at fixed filling. Circles: polaron Ansatz~\eqref{eq:2p2h}  with $\bQ_P=0$, squares: dimeron Ansatz~\eqref{eq:dimeron} with $\mathbf Q_M=0$. Line: exact $\eta$-dimer energy. (i) $\rho_\uparrow=0.02$: energies are measured from $E_{2b}-\epsilon_F$, where $E_{2b}$ is the two-body bound-state energy on the 
lattice. The dimeron crosses below the polaron, and the $\eta$ dimer below the dimeron at stronger coupling. (ii) $\rho_\uparrow=0.106$ and (iii) $\rho_\uparrow=0.30$: energies are measured from the $\eta$-dimer energy. The polaron crosses the $\eta$ dimer in (ii), and remains below it at every coupling in (iii).}
\label{fig:scenarios}
\end{figure}

In this Letter we show that on the lattice the polaron competes not only with the dimeron of the continuum but also with a different bound state: 
Yang's $\eta$ pair~\cite{Yang1989}, which we call the $\eta$ dimer, 
a single doublon delocalized with momentum $\bPi=(\pi,\pi)$, which is an exact eigenstate at every coupling.
 Under a partial particle--hole transformation it becomes the Nagaoka ferromagnet of the repulsive Hubbard model~\cite{Nagaoka1966,Thouless1965,Tasaki1989}, so that the competition between the polaron and the $\eta$ dimer is the single-spin-flip stability problem of Nagaoka ferromagnetism, studied for decades~\cite{SKA1990,Hanisch1993,Wurth1995,Wurth1996,Hanisch1997,BeccaSorella2001,Carleo2011,Liu2012}. Using variational Ans\"atze with up to two particle--hole excitations (2p2h) in 
  the thermodynamic limit, benchmarked against sign-problem-free  determinant DiagMC (PDet algorithm~\cite{Vanhoucke2020,Pascual2025}), 
 we obtain the full ground-state phase diagram (Fig.~\ref{fig:phase}). 
 It reveals three scenarios as the attraction increases (Fig.~\ref{fig:scenarios}): 
 (i) at very low filling, $\rho_\uparrow\lesssim0.045$, the ground state changes from the polaron to the dimeron and then to the $\eta$ dimer, (ii) at intermediate filling, it changes directly from the polaron to the $\eta$ dimer, and (iii) above $\rho_\uparrow\approx0.248$, it remains the polaron at every coupling.
On the lattice,  the dimeron is thus  itself replaced by the $\eta$ dimer at strong coupling, and the dimeron region ends at a triple point. 

In the repulsive language, the two boundaries of the $\eta$-dimer region correspond to the two known instability mechanisms of the Nagaoka state, 
an instability driven by the antiferromagnetic exchange $J=4t^2/|U|$ at low hole density~\cite{Nagaoka1966,Richmond1969,SKA1990} and 
a single-particle (Stoner) instability at higher density~\cite{SKA1990,vonderLinden1991}. 
To our knowledge, the triple point where they meet has not been identified. Since the $\eta$ dimer is an exact eigenstate, its phase boundary gives a rigorous criterion for the instability of the Nagaoka state against a single spin flip, which improves on the best previous one~\cite{Wurth1996} at all fillings. 
At low filling the critical coupling $|U_c|$ of this boundary grows as $(4t/\pi)\ln(1/\rho_\uparrow)/\rho_\uparrow$, a logarithmic enhancement absent from previous criteria~\cite{Hanisch1993,Wurth1996,Hanisch1997}.

\emph{Model and the $\eta$ dimer.}---We consider the attractive Hubbard model on the square lattice with $N=L^2$ sites, lattice spacing one, and periodic boundary conditions,
\begin{equation}
\hat H=-t\sum_{\langle\br\br'\rangle,\sigma}\left(\cd_{\br\sigma}\ca_{\br'\sigma}+\text{h.c.}\right)+U\sum_\br\hat n_{\br\uparrow}\hat n_{\br\downarrow}\; ,
\label{eq:H}
\end{equation}
where $\cd_{\br\sigma}$ ($\ca_{\br\sigma}$) creates (annihilates) a fermion of spin $\sigma=\uparrow,\downarrow$ on site $\br$, $\hat n_{\br\sigma}=\cd_{\br\sigma}\ca_{\br\sigma}$, hopping amplitude $t$ and on-site attraction $U<0$.
We consider $N_\uparrow$ majority fermions, with filling $\rho_\uparrow=N_\uparrow/N$, and a single spin-down fermion.
The dispersion is $\epsilon_\bk=-2t(\cos k_x+\cos k_y)$. 
The filled Fermi sea state of $N_\uparrow$ fermions is written as $\FS{N_\uparrow}$, and has total energy $\EFS{N_\uparrow}$.
Quasiparticle energies are defined relative to the energy of the Fermi sea, $E_{\rm QP}=E-\EFS{N_\uparrow}$.

On a bipartite lattice, $\epsilon_{\bPi-\bk}=-\epsilon_\bk$. The $\eta$-pair operator~\cite{Yang1989,Zhang1990},
\begin{equation}
\hat\eta^\dagger=\sum_\br e^{i\bPi\cdot\br}\,\cd_{\br\uparrow}\cd_{\br\downarrow}=\sum_\bk\cd_{\bk\uparrow}\cd_{\bPi-\bk,\downarrow} \; ,
\label{eq:eta}
\end{equation}
therefore commutes with the kinetic energy, and, acting on a Fermi sea of spin-up fermions, it creates exactly one doubly occupied site. 
The state
\begin{equation}
|\eta\rangle=\frac{1}{\sqrt{N-N_\uparrow+1}}\,\hat\eta^\dagger\,\FS{N_\uparrow-1}
\label{eq:etastate}
\end{equation}
is an eigenstate of the Hamiltonian~\eqref{eq:H} at every $U$, with energy $E^{(\eta)}$ so that the $\eta$-dimer quasiparticle energy is
\begin{equation}
E^{(\eta)}_{\rm QP}= E^{(\eta)} -\EFS{N_\uparrow} =  -\epsilon_F+U \; ,
\label{eq:Eeta}
\end{equation}
with $\epsilon_F\equiv\EFS{N_\uparrow}-\EFS{N_\uparrow-1}$.
It is a coherent superposition of the doublon on all sites with a staggered phase or, equivalently, of all pairs $(\bk,\bPi-\bk)$ with $\bk$ outside the Fermi sea $\FS{N_\uparrow-1}$.
Each component has zero kinetic energy, $\epsilon_\bk+\epsilon_{\bPi-\bk}=0$, so the pair costs no kinetic energy and leaves the remaining $N_\uparrow-1$ fermions unperturbed. Its quasiparticle residue is $Z=1/(N-N_\uparrow+1)\to0$ in the thermodynamic limit, so that, like the dimeron, the $\eta$ dimer is a bosonic bound state. 
Note that Eq.~\eqref{eq:Eeta} gives an exact upper bound $E_{\rm QP}\le U-\epsilon_F$ on the ground-state energy at every $U$ and $\rho_\uparrow$.

\emph{Mapping onto the Nagaoka problem.}---
The particle--hole transformation $\da_{\br\uparrow}=e^{i\bPi\cdot\br}\cd_{\br\uparrow}$, $\da_{\br\downarrow}=\ca_{\br\downarrow}$ leaves the hopping invariant and gives
\begin{equation}
\hat H=U\hat N_\downarrow+\hat H_{\rm rep}(|U|) \; ,
\label{eq:map}
\end{equation}
where $\hat H_{\rm rep}$ is the repulsive Hubbard model. 
The Fermi sea $\FS{N_\uparrow-1}$ becomes the fully polarized state $|{\rm FM}\rangle$ of the repulsive model, the Fermi sea of $N-N_\uparrow+1$ spin-up fermions, with total spin $S=S^z=S_{\rm max}\equiv(N-N_\uparrow+1)/2$.
Eq.~\eqref{eq:eta} becomes $\hat\eta^\dagger=-\hat S^-$, with $\hat S^-=\sum_\br\dd_{\br\downarrow}\da_{\br\uparrow}$ the total spin-lowering operator of the repulsive model.
Throughout this Letter, $S$ denotes the total spin of the repulsive model.
 Hence
\begin{equation}
|\eta\rangle\propto\hat S^-|{\rm FM}\rangle ,
\label{eq:goldstone}
\end{equation}
which is a member of the multiplet of the Nagaoka ferromagnet, namely  the Nagaoka state rotated by a uniform spin flip.
In this single-spin-flip sector, every eigenstate of the repulsive model either has total spin $S_{\rm max}$ and is of the form $\hat S^-|\Phi\rangle$, 
with $|\Phi\rangle$ a fully polarized eigenstate (i.e.\ a Slater determinant of spin-up plane waves), 
or has total spin $S_{\rm max}-1$. The polaron and the dimeron map onto states with $S=S_{\rm max}-1$.
The $\eta$ dimer is thus the ground state exactly when no state with $S=S_{\rm max}-1$ lies below the fully polarized ferromagnet, at hole density $\delta=(N_\uparrow-1)/N\to\rho_\uparrow$ and repulsion $|U|$. In the language of the Nagaoka problem, the ferromagnet is then stable against a single spin flip.
This correspondence between $\eta$ pairing and Nagaoka ferromagnetism was established in Ref.~\cite{SinghScalettar1991}.
Here we exploit the exactness of $|\eta\rangle$: variational polaron and dimeron energies below it give a rigorous criterion for the instability of the Nagaoka state.
Moreover, for $N_\uparrow=2$ (one hole in the repulsive model), Nagaoka's theorem~\cite{Nagaoka1966,Thouless1965,Tasaki1989} ensures that $|\eta\rangle$ is the exact ground state as $U\to-\infty$.
Table~\ref{tab:dictionary} summarizes the correspondence between the states of the two models.

\begin{table}[t]
\caption{Single-impurity states of the attractive model and their counterparts in the repulsive model at hole density $\delta=\rho_\uparrow$, related by the partial particle--hole transformation of Eq.~\eqref{eq:map}. Relative to the ferromagnet, each state is a single spin flip, i.e.\ a spin-up hole and a spin-down fermion, either unbound, as in a Stoner excitation, or bound in a magnon~\cite{SKA1990}. A dimer of pair momentum $\mathbf Q_M$ [Eq.~\eqref{eq:dimeron}] has magnon momentum $\mathbf q=\mathbf Q_M+\bPi$. The $\mathbf q=0$ magnon is the Goldstone mode of the ferromagnet, i.e.\ its uniform rotation [Eq.~\eqref{eq:goldstone}], and is an exact, undressed eigenstate. $S$ is the total spin of the repulsive model.}
\label{tab:dictionary}
\begin{ruledtabular}
\begin{tabular}{llc}
attractive ($U<0$) & repulsive ($U>0$) & $S$\\
\hline
polaron & dressed Stoner excitation & $S_{\rm max}-1$\\
dimeron & dressed magnon, $\mathbf q=\bPi$ & $S_{\rm max}-1$\\
$\eta$ dimer & magnon, $\mathbf q=0$ & $S_{\rm max}$\\
\end{tabular}
\end{ruledtabular}
\end{table}

\emph{Methods.}---To establish the phase diagram, we use variational Ans\"atze with up to two particle--hole excitations of the Fermi sea. For the polaron,
\begin{multline}
|P(\bQ_P)\rangle=\Big[\alpha_0\cd_{\bQ_P\downarrow}
+\frac1N\sum_{\bk\bq}{}^{'}\alpha_{\bk\bq}\,\cd_{\bQ_P+\bq-\bk,\downarrow}\cd_{\bk\uparrow}\ca_{\bq\uparrow}\\
+\frac{1}{4N^2}\sum_{\bk\bk'\bq\bq'}{}^{'}\alpha_{\bk\bk'\bq\bq'}\,\cd_{\bQ_P+\bq+\bq'-\bk-\bk',\downarrow}\\
\times\cd_{\bk\uparrow}\cd_{\bk'\uparrow}\ca_{\bq'\uparrow}\ca_{\bq\uparrow}\Big]\FS{N_\uparrow},
\label{eq:2p2h}
\end{multline}
where $N_\uparrow$ is chosen such that $\FS{N_\uparrow}$ is a non-degenerate closed shell with zero total quasi-momentum. For the dimeron, whose bare pair is built on the Fermi sea with one fermion removed, $|{\rm FS}^{(\mathbf Q_{\rm FS})}_{N_\uparrow-1}\rangle=\ca_{-\mathbf Q_{\rm FS},\uparrow}\FS{N_\uparrow}$ with $\mathbf Q_{\rm FS}$ on the Fermi surface, and dressed by up to two particle--hole excitations,
\begin{multline}
|M(\mathbf Q_M)\rangle=\Big[\frac{1}{\sqrt N}\sum_\bk{}^{'}\xi_\bk\,\cd_{\mathbf Q_M-\bk,\downarrow}\cd_{\bk\uparrow}\\
+\frac{1}{2N^{3/2}}\sum_{\bk\bk'\bq}{}^{'}\xi_{\bk\bk'\bq}\,\cd_{\mathbf Q_M+\bq-\bk-\bk',\downarrow}\cd_{\bk\uparrow}\cd_{\bk'\uparrow}\ca_{\bq\uparrow}\\
+\frac{1}{12N^{5/2}}\sum_{\bk\bk'\bk''\bq\bq'}{}^{'}\xi_{\bk\bk'\bk''\bq\bq'}\,\cd_{\mathbf Q_M+\bq+\bq'-\bk-\bk'-\bk'',\downarrow}\\
\times\cd_{\bk\uparrow}\cd_{\bk'\uparrow}\cd_{\bk''\uparrow}\ca_{\bq'\uparrow}\ca_{\bq\uparrow}\Big]|{\rm FS}^{(\mathbf Q_{\rm FS})}_{N_\uparrow-1}\rangle.
\label{eq:dimeron}
\end{multline}
The primes restrict $\bk,\bk',\bk''$ to levels outside and $\bq,\bq'$ to levels inside the respective Fermi sea~\cite{Punk2009,Combescot2008,Pascual2025}. The total momentum is $\bQ_P$ for Eq.~\eqref{eq:2p2h} and $\mathbf Q_M+\mathbf Q_{\rm FS}$ for Eq.~\eqref{eq:dimeron}. We solve the variational equations both on finite lattices and directly in the thermodynamic limit, where the sums become integrals over the Brillouin zone. 
The two Ans\"atze overlap: at finite quasi-momentum, Ansatz~\eqref{eq:2p2h} can also describe bound states, as it does in the continuum~\cite{Cui2020,Peng2021}, and at $\bQ_P=\bPi+\mathbf Q_{\rm FS}$ it contains $|\eta\rangle$ exactly, whose known energy provides a check of the numerics. We therefore identify the states by their total momentum. 
Scanning all momenta, we find that wherever the polaron or the dimeron is the ground state, its energy is lowest at $\bQ_P=0$ or $\mathbf Q_M=0$, respectively, while the $\eta$ dimer has $\mathbf Q_M=\bPi$.
The three states thus lie in different momentum sectors, and the transitions between them are sharp level crossings.

Wherever the 2p2h polaron or dimeron energy lies below $U-\epsilon_F$, the Nagaoka state is  unstable against a single spin flip. Since the interaction in $\hat H_{\rm rep}$ is non-negative, $E_{\rm QP}-U$ is non-decreasing in $|U|$, whereas $E^{(\eta)}_{\rm QP}-U=-\epsilon_F$ is constant. The $\eta$ dimer is thus the ground state only above a critical coupling $|U_c(\rho_\uparrow)|$, for which the boundary in Fig.~\ref{fig:phase} is a rigorous lower bound within the numerical accuracy of the thermodynamic-limit solution. 
The polaron--dimeron boundary, by contrast, compares two variational energies.

\begin{figure}[t]
\includegraphics[width=\columnwidth]{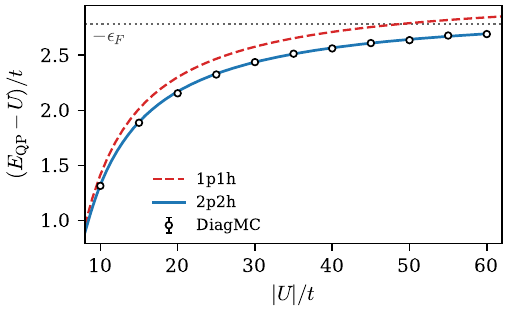}
\caption{Polaron energy at $\rho_\uparrow=0.105$ from PDet DiagMC (circles), compared with variational Ansatz with one (dashed) and two (solid) particle--hole excitations. The dotted line is the exact $\eta$-dimer energy, $E^{(\eta)}_{\rm QP}-U=-\epsilon_F$.  }
\label{fig:mc}
\end{figure}

To assess the accuracy of the variational Ansatz, we compare with the PDet DiagMC method~\cite{Vanhoucke2020,Pascual2025} for the bare-interaction expansion of the impurity Green's function. Without a chemical-potential shift, the algorithm turns out to be sign-problem free~\cite{Pascual2025}, which allows us to reach very high diagram order and thus very high $|U|/t$. 
For the continuum polaron at unitarity in three dimensions, the 2p2h energy agrees with PDet to four significant digits~\cite{Vanhoucke2020}, and the 2p2h variational Ansatz is also highly accurate in two dimensions~\cite{Parish2013,Vlietinck2014}. 
We therefore expect the low-filling part of the phase diagram to be highly accurate, and we test the variational Ansatz where lattice effects are stronger, at the higher filling $\rho_\uparrow=0.105$ and up to $|U|=60t$. 
 Figure~\ref{fig:mc} shows the polaron energy obtained with PDet (for $L=74$ and $N_\uparrow=577$, where finite-size effects are small) and with the thermodynamic-limit variational Ansatz, providing a stringent test of the 2p2h wave function at strong coupling.  
 The 2p2h energy agrees with PDet to within $0.017t$ over the whole range, and to within 1.5 standard deviations of PDet for $|U|\ge30t$. Where the two differ, PDet lies lower, as expected for a variational upper bound.
 The single particle--hole result, by contrast, deviates by up to $\approx0.15t$ and crosses the $\eta$-dimer energy already at $|U|\approx49t$, about half the 2p2h value.

\emph{Phase diagram.}---Figure~\ref{fig:scenarios} shows the energies along three vertical cuts through Fig.~\ref{fig:phase}, one for each scenario.
At weak coupling the ground state is always the polaron. At low filling, the polaron is replaced by the dimeron as the coupling increases, as in the continuum, at $|U|\approx11.7t$ for $\rho_\uparrow=0.02$.
As $\rho_\uparrow\to0$ this boundary approaches the continuum polaron--molecule transition~\cite{Parish2013,Vlietinck2014}, $E_b=C(\epsilon_F+4t)$ with $E_b=-8t-E_{2b}$ the two-body binding energy and $C\approx10$--$14$, so that $|U_c|$ vanishes as $8\pi t/\ln(1/\rho_\uparrow)$.
The dimeron region, however, ends at a triple point at 
$\rho_\uparrow\approx0.045$ and $|U|\approx146t$. For larger filling, no dimeron at rest is ever the ground state. Instead, the ground state changes directly from the polaron to the $\eta$ dimer, at $|U|\gtrsim92t$ near $\rho_\uparrow\approx0.106$. 
The $\eta$-dimer region ends at $\rho_\uparrow\approx0.2476$, where the boundary diverges, as it also does for $\rho_\uparrow\to0$ (see below).

In the repulsive language, the $|U|\to\infty$ limit of the boundary proves that the Nagaoka state is unstable against a single spin flip for 
$\delta>0.2476$, compared with $\delta>0.2514$ from
the variational wave function of WUM~\cite{Wurth1996} and $\delta>0.49$ from the Shastry--Krishnamurthy--Anderson (SKA) state~\cite{SKA1990}. 
At $U\to-\infty$ we find that this boundary is set by the polaron, not by the bound spin wave at $(0,\pi)$ proposed in Refs.~\cite{Wurth1995,Wurth1996}, which in our language is the dimeron at $\mathbf Q_M=(\pi,0)$. At the 2p2h level the latter lies $(0.015\pm0.005)t$ above the polaron at $\rho_\uparrow=0.25$ and $0.30$.

At finite $U$ our boundary lies above the WUM curve at all fillings, with a minimum of $92t$ instead of $77.7t$, about 20\% higher, and up to 40\% higher at $\rho_\uparrow=0.02$.
To our knowledge this is the most stringent rigorous criterion for the instability of the Nagaoka state on the square lattice. 
Since stability against all spin flips requires stability against one, it also bounds the region of saturated ferromagnetism, which numerical studies place at $\delta\lesssim0.2$~\cite{Liu2012}
to $0.25$~\cite{Carleo2011}, consistent with our bound.

\emph{Competition of kinetic energies.}---At strong coupling all three states contain exactly one doublon, so they have the same interaction energy $U$, and they differ only in their kinetic energy.
The doublon of the dimeron gains kinetic energy by pair hopping, of order $t^2/|U|$, which requires empty neighboring sites. 
The polaron propagates by partner exchange: the spin-down fermion hops, at first order in $t$, onto a neighboring site occupied by a spin-up fermion. 
Through repeated exchanges, the spin-down fermion forms a doublon with every fermion of the sea in turn, so that all $N_\uparrow$ spin-up fermions remain part of the Fermi sea. This is why the polaron keeps a finite residue $Z$. In both dimers, by contrast, the Fermi sea contains only $N_\uparrow-1$ fermions, and the remaining spin-up fermion is bound in the doublon, which gives $Z\propto1/N$.
The $\eta$ dimer has no kinetic energy of its own: the doublon carries none, and the remaining Fermi sea is left untouched. Its only cost is the spin-up fermion taken from the Fermi level, which gives $E^{(\eta)}_{\rm QP}-U=-\epsilon_F$.

Partner exchange is captured by the strong-coupling polaron state
\begin{equation}
|P\rangle=\frac{1}{\sqrt{N_\uparrow}}\sum_\br\hat n_{\br\uparrow}\cd_{\br\downarrow}\FS{N_\uparrow},
\label{eq:Pstate}
\end{equation}
in which the spin-down fermion is placed only on sites occupied by the Fermi sea, so that it always forms a doublon. This state is contained in Ansatz~\eqref{eq:2p2h}, with $\bQ_P=0$, $\alpha_0=\sqrt{\rho_\uparrow}$, $\alpha_{\bk\bq}=1/\sqrt{\rho_\uparrow}$ and no two-particle--hole term. It has residue $Z=\rho_\uparrow$, and in the repulsive language it coincides with the single-spin-flip state of SKA~\cite{SKA1990}. Its energy is
\begin{equation}
E^{(P)}_{\rm QP}-U=-\bar\epsilon-zt\rho_\uparrow+\frac{\rho_\uparrow\bar\epsilon^{\,2}}{zt},
\label{eq:EP}
\end{equation}
with $z=4$ and $\bar\epsilon=\EFS{N_\uparrow}/N_\uparrow$.
The three terms are the average energy of the spin-up fermion bound in the doublon, the hopping of the spin-down fermion onto occupied neighboring sites, and the reduction of the number of occupied neighbors by Pauli exclusion among the spin-up fermions~\cite{SKA1990}.
At low filling the last two terms cancel to order $\rho_\uparrow$ and $-\bar\epsilon$ exceeds $-\epsilon_F$, so the $\eta$ dimer is lower in energy, whereas at higher filling the hopping dominates and Eq.~\eqref{eq:EP} crosses Eq.~\eqref{eq:Eeta} at $\rho_\uparrow=0.491$, a point that particle--hole dressing moves down to $\rho_\uparrow\approx0.248$.

Below the triple point, $\rho_\uparrow\lesssim0.045$, the $\eta$ dimer competes instead with the dimeron. The dimeron gains the pair-hopping energy $16t^2/|U|$, but its doublon hinders the motion of the remaining spin-up fermions, which costs energy, whereas the $\eta$ dimer leaves them untouched. In two dimensions this cost per fermion vanishes logarithmically at low density, so that $|U_c|$ grows as $(4t/\pi)\ln(1/\rho_\uparrow)/\rho_\uparrow$, consistent with our 2p2h boundary. In the repulsive language this is the exchange-driven instability of the Nagaoka state at low density~\cite{Nagaoka1966,SKA1990}.

Near full filling, Eq.~\eqref{eq:map} maps the problem onto a single spin-down fermion in a dilute spin-up Fermi gas of the repulsive model, with Fermi energy $4t-\epsilon_F$ measured from the band bottom. The energy $E_{\rm QP}-U$ is non-decreasing in $|U|$ and therefore bounded by its hard-core value. 
In two dimensions even a hard-core repulsion scatters weakly at low energy, with a $T$ matrix $8\pi t/\ln(t/\varepsilon)$ at energy $\varepsilon$ above the two-particle threshold $2\epsilon_0=-8t$.
To first order in the density, $E_{\rm QP}-U+4t$ is then $(1-\rho_\uparrow)$ times this $T$ matrix at $\varepsilon=4t-\epsilon_F$, i.e.\ $2(4t-\epsilon_F)/\ln[t/(4t-\epsilon_F)]$.
The $\eta$ dimer, in contrast, costs the full $4t-\epsilon_F$.
The ratio of the two energies, $2/\ln[t/(4t-\epsilon_F)] \equiv \nu T$ with $\nu=1/(4\pi t)$ the density of states near the band bottom, is the dimensionless coupling of the problem.
Since it vanishes as $\rho_\uparrow\to1$ at every $U$, no bound state, neither the $\eta$ dimer nor the dimeron, forms below the polaron.

The results extend beyond a single impurity. The states $(\hat\eta^\dagger)^{N_\downarrow}\FS{N_\uparrow-N_\downarrow}$ are exact eigenstates that map onto the Nagaoka multiplet at hole density $\rho_\uparrow-\rho_\downarrow$. 
Wherever the fully polarized ferromagnet, with total spin $S=(N-N_\uparrow+N_\downarrow)/2$, is the ground state of the repulsive model, the spin-imbalanced attractive gas has thus an $\eta$-paired ground state
with off-diagonal long-range order at momentum $\bPi$~\cite{Yang1989,SinghScalettar1991,SteinOppermann1992}. 
A fully polarized ground state is in particular stable against a single spin flip. A necessary (but not sufficient) condition for the $\eta$-paired ground state is therefore that the $\eta$ dimer be the single-impurity ground state at the same hole density, i.e.\ at filling $\rho_\uparrow-\rho_\downarrow$ in Fig.~\ref{fig:phase}.

\emph{Conclusion.}---The lattice Fermi polaron has three ground states. The dimeron of the continuum survives only at low filling, $\rho_\uparrow\lesssim0.045$, and its region ends at a triple point. At strong coupling it is replaced by the $\eta$ dimer, an exact eigenstate that maps onto the uniformly rotated Nagaoka ferromagnet, and above $\rho_\uparrow\approx0.248$ the polaron is the ground state at all couplings. Because the $\eta$ dimer is exact, our 2p2h results, benchmarked against diagrammatic Monte Carlo, yield a rigorous instability criterion for Nagaoka ferromagnetism that is sharper than the best previous one at all fillings.
At low filling the critical coupling exceeds the inverse-filling scaling of previous criteria by a logarithmic factor.
Quantum-gas experiments on the attractive Hubbard model reach couplings up to $|U|\approx16t$~\cite{Hartke2023}, beyond the polaron--dimeron transition at low filling, which is thus within experimental reach, complementary to the imaging of Nagaoka polarons in the repulsive model~\cite{Lebrat2024,Prichard2024}.
The $\eta$ region requires $|U|\gtrsim92t$. Wherever Nagaoka ferromagnetism is stable, the exact mapping moreover implies an $\eta$-paired ground state of the spin-imbalanced gas, for which our phase boundary provides a necessary condition.
Natural extensions are the cubic lattice, frustrated lattices, where the $\eta$ dimer is no longer exact, and a finite density of impurities beyond the necessary condition derived here.

\begin{acknowledgments}
We thank Jordi Boronat, Boris Svistunov and F\'elix Werner for valuable discussions. This work was supported by the French Agence Nationale de la Recherche (ANR) under grant ANR-21-CE30-0033 (project LODIS). 
The authors used Claude (Anthropic, Claude Opus 5.5) to edit the text, to design code for the variational calculations, to optimize some of our existing codes, and to check analytical arguments. The authors directed this use, verified all derivations and numerical results independently, and take full responsibility for the content of the paper.
\end{acknowledgments}

\bibliography{prl_refs}

\end{document}